\documentclass[pra,aps,twocolumn,tightlines,showpacs]{revtex4}

\usepackage[]{times,amsmath,epsfig,graphicx}

\begin{document}

\title{Transient Regime of Kerr Frequency Comb Formation}

\author{A. A. Savchenkov, A. B. Matsko, W. Liang, V. S. Ilchenko, D. Seidel, and L. Maleki}

\affiliation{OEwaves Inc., 465 N. Halstead St. Ste. 140, Pasadena, CA 91107}

\begin{abstract}
Temporal growth of an optical Kerr frequency comb generated in a microresonator is studied both
experimentally and numerically. We find that the comb emerges from vacuum fluctuations of the
electromagnetic field on timescales significantly exceeding the ringdown time of the resonator modes.
The frequency harmonics of the comb spread starting from the optically pumped mode if the
microresonator is characterized with anomalous group velocity dispersion. The harmonics have
different growth rates resulting from sequential four-wave mixing process that explains intrinsic
modelocking of the comb.
\end{abstract}


\maketitle

Kerr combs excited in nonlinear optical microresonators hold promise as chip scale generators of
octave spanning optical frequency combs with unique characteristics \cite{kippenberg11s}. They
also reveal a rich, and as yet not well understood, variety of nonlinear dynamical phenomena.
Different regimes of comb generation have been observed and several theoretical models developed.
While the mainfocus of the research in these studies is related to understanding  the spectral properties of the
Kerr comb, less work has been devoted to its time domain behavior. Several experimental
\cite{arcizet09ch,foster11oe,ferdous11np,papp11arch} and theoretical
\cite{agha09oe,chembo10prl,chembo10pra,matsko11ol} studies of mode locking regimes of this nonlinear
process have been published very recently, but the problem of temporal growth of comb harmonics
was addressed only theoretically \cite{chembo10prl,chembo10pra}. In this Letter we report on
experimental study of  transient dynamics of Kerr frequency combs and also propose a theoretical
explanation for the observed results. We also suggest a numerical model, backed by the experiment,
that predicts a different dynamics for Kerr comb growth compared with earlier theoretical
predictions. Our research clearly explains why harmonics of the comb are modelocked.

Kerr combs are generated from  electromagnetic vacuum fluctuations due to modulation instability
of a continuous wave (cw) light confined in an externally pumped nonlinear dispersive resonator. When
the power of the cw optical pump that is nearly resonant with one of the modes of the resonator
exceeds a certain threshold, the cw field inside the resonator becomes unstable, and multiple
frequency harmonics are generated in the modes. The harmonics are equally spaced due to energy and
photon number conservation laws, imposed by four-wave mixing process (FWM), which is responsible
for  comb generation \cite{delhaye11prl,savchenkov11np}.

The growth of the comb is not instantaneous. It was found \cite{chembo10prl,chembo10pra} that
formation of a fully developed comb can take up to a hundred ring-down periods of the resonator mode.
The goal of the present contribution is to measure this time interval directly. We performed a
numerical simulation and found that the DC power of light exiting the resonator depends on the degree
of comb formation. The effective intrinsic quality factor of the pumped optical mode suddenly drops
long after the steady state amplitude of the circulating light in the mode is reached. The reduction
of the intrinsic quality factor changes the attenuation of the pump mode since the balance between
intrinsic loss and coupling loss in the resonator changes. Hence, the time delay between the
moment the light enters the pump mode and the moment when the frequency comb is generated can be
directly measured by detecting the light escaping the resonator. It is also possible to track the
time dependence of the power of the RF signal generated by the comb to measure the delay. We
performed such a measurement with a Kerr frequency comb produced in a high-Q CaF$_2$ whispering
gallery mode (WGM) resonator, and experimentally confirmed the theoretical prediction of
\cite{chembo10prl,chembo10pra} as well as the result of our numerical simulations. In what follows we
describe our simulations and the experiment in detail.

To simulate the transient regime of the Kerr comb we use the theoretical model developed in
\cite{matsko05pra,chembo10pra}. We introduce an interaction Hamiltonian
\begin{equation} \label{v}
\hat V= -\frac{\hbar g}{2} (\hat e^\dag)^2 \hat e^2,
\end{equation}
where
\begin{equation} \label{g}
g=\frac{\hbar \omega_0^2 c n_2}{{\cal V}n_0^2}
\end{equation}
is the coupling parameter obtained under assumption of complete space overlap of the resonator modes,
$\omega_0$ is the value of the optical frequency of the externally pumped mode, $c$ is the speed of
light in vacuum, $n_2$ is the cubic nonlinearity of the material, ${\cal V}$ is the effective
geometrical volume occupied by the optical modes in the resonator, and $n_0$ is the linear index of
refraction of the resonator host material. The operator $\hat e$ is given by the sum of annihilation
operators of the electromagnetic field for 41 interacting resonator modes that we took into
consideration:
\begin{equation}
\hat e= \sum \limits_{j=1}^{41} \hat a_j.
\end{equation}
The external cw pump is applied to the central mode of the group, so the Kerr frequency comb is
expected to have twenty red- and twenty blue-detuned harmonics with respect to the frequency of the
pumped mode. We take into account only the second order frequency dispersion that is recalculated for
the frequency of the modes.

Equations describing the evolution of the field in the resonator modes are generated using 
Hamiltonian (\ref{v}) and input-output formalism developed for ring resonators \cite{yariv02ptl}
\begin{equation} \label{set}
\dot{\hat a}_j=-(\gamma_0+i\omega_j)\hat a_j+ \frac{i}{\hbar} [\hat V,\hat a_j]+F_0 e^{-i\omega t}
\delta_{21,j},
\end{equation}
where $\delta_{21,j}$ is the Kronecker's delta; $\gamma_0=\gamma_{0c}+\gamma_{0i}$ is the half width
at the half maximum for the optical modes, assumed to be the same for the all modes involved; and
$\gamma_{0c}$ and $\gamma_{0i}$ stand for coupling and intrinsic loss. The external optical pumping
is given by
\begin{equation}
F_0= \sqrt{\frac{ 2 \gamma_{0c} P}{\hbar \omega_0}},
\end{equation}
where $P$ is the value of the cw pump light. We neglect the quantum effects and do not take into
account corresponding Langevin noise terms.

The set (\ref{set}) should be supplied with an equation describing the light leaving the resonator.
Assuming the pump power does not depend on time, the relative amplitude of the output field is given
by \cite{yariv02ptl}
\begin{equation}\label{output}
\frac{\hat e_{out}}{\hat e_{in}}=  \sqrt{1-2\gamma_{0c} \tau_0}-\sqrt{1-2\gamma_{0i} \tau_0}
\frac{2\gamma_{0c}}{\gamma_{0c}+\gamma_{0i}} \frac{\gamma_0 \hat e}{F_0},
\end{equation}
where $\tau_0 = 2 \pi R n_0/c$ is the light round trip time for the resonator, and $R$ is the radius
of the resonator. It is assumed that $\hat e(t-\tau_0) \approx \hat e(t)$.
\begin{figure}[ht]
  \centering
  \includegraphics[width=8.5cm]{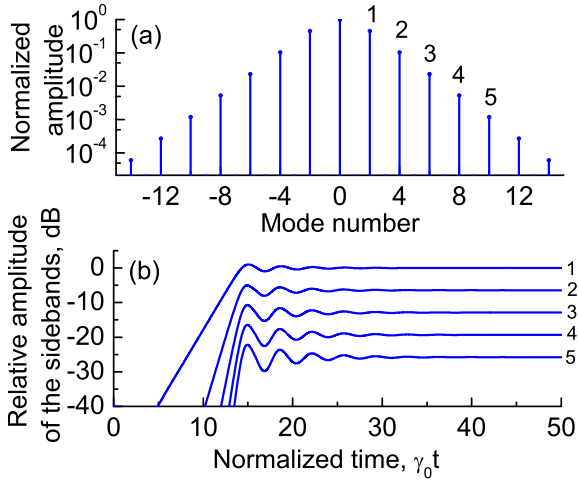}
\caption{ (a) Optical frequency comb generated in the WGM resonator for  selected numerical
parameters of the system. (b) Time dependence of the normalized amplitudes of the first five
harmonics of the optical frequency comb generated in the WGM resonator characterized with
$(g/\gamma_0)^{1/2}=5\times10^{-5}$.  } \label{fig1}
\end{figure}

We solved Eqs.~(\ref{set}) and (\ref{output}) numerically taking into account the interaction of 41
optical modes. Since the number of modes in our numerical simulation is limited, we  selected a
resonator with large GVD to be sure that the spectral boundary conditions do not impact the
nonlinear process. We assumed $2\omega_{21}-  \omega_{22}-  \omega_{20} = -\gamma_{0}$,
$(F_0/\gamma_0)(g/\gamma_0)^{1/2}=1.5$, and $\omega=\omega_{21}-1.7 \gamma_0$, and found that the
resonator generates the optical frequency comb with spectrum shown in Fig.~(\ref{fig1}a). The comb
has a frequency harmonic in each second optical mode (for the selected values of the parameters of
the system), which sometimes is observed experimentally (see Fig.~5 in \cite{maleki10ifcs}). The
different regimes of Kerr comb generation will be studied elsewhere, while in this work we focus on
the transient processes.

\begin{figure}[ht]
  \centering
  \includegraphics[width=8.5cm]{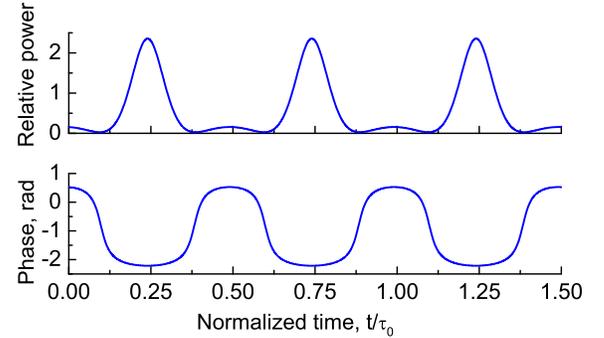}
\caption{ Relative power and phase for the optical pulses leaving the resonator. The shape of these pulses is similar to the shape of the pulses reported in \cite{arcizet09ch}.} \label{fig2}
\end{figure}

We performed a numerical simulation of the growth of the first five harmonics of the frequency comb
Fig.~(\ref{fig1}b) revealing several important features of comb generation. The sidebands grow
exponentially with different growth rates. The first harmonic ((1) in Fig.~\ref{fig1}) has the
slowest growth rate of $\gamma_0$, the second harmonic ((2) in Fig.~\ref{fig1}) has twice faster
growth rate, $2\gamma_0$, the third harmonic -- $3\gamma_0$. The first harmonic starts to grow much
earlier than  others, the second harmonic starts before the third, etc. Such a temporal behavior
shows that the particular realization of the Kerr comb is initiated by hyper-parametric oscillation
\cite{matsko05pra} that involves only two optical sidebands closest in frequency to the pump. The
next order of optical harmonics is generated in the stimulated comb due to FWM of the already
generated sidebands and the pump light \cite{maleki10ifcs}. This stimulated process does not have a
threshold. In other words, the threshold of comb generation coincides with the threshold of the
hyper-parametric oscillation. {\em The lowest order harmonics also determine the phase as well as the
frequency of the rest of the harmonics. Such a comb is always phase locked and optical pulses are
formed in the resonator } (Fig.~\ref{fig2}).

\begin{figure}[ht]
  \centering
  \includegraphics[width=8.5cm]{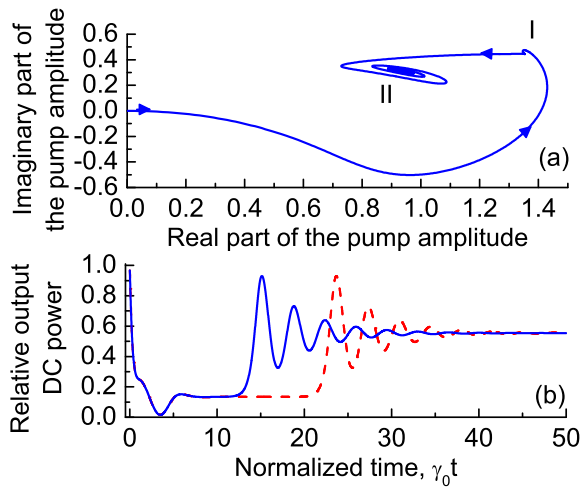}
\caption{(a) Transient behavior of the normalized amplitude of the electromagnetic field of the externally pumped optical mode, $\langle \hat a_{21}(t) \rangle$. Since the mode is initially empty, its amplitude increases and reaches the attractor (I) approximately during resonator's ring down time, after the cw pump light is on. Then, after a certain period of time equal to the time interval needed for the comb to grow, the amplitude drops to a certain level and changes its phase  approaching another attractor, (II). (b) The transient behavior of the light exiting the resonator is calculated for slightly overcoupled modes, $\gamma_{0c}=1.5\gamma_{0i}$, and for different values of nonlinearity $g$. The oscillator reaches its steady state faster for larger nonlinearity ($(g/\gamma_0)^{1/2}=5\times10^{-5}$, solid blue line), and slower for smaller nonlinearity ($(g/\gamma_0)^{1/2}= 10^{-8}$, dashed red line).} \label{fig3}
\end{figure}

Another important observation is related to the temporal behavior of the pump light. Since the growth
of the comb harmonics is not instantaneous, the pump light is not influenced by the comb growth
initially. The pump power is impacted only after the frequency harmonics approach their saturation
values. This behavior is clearly seen at the phase diagram (Fig.~\ref{fig3}a), where the pump mode
has two attractors. The first one ((I) in Fig.~\ref{fig3}a) corresponds to the steady state solution
for for the pump light in the nonlinear resonator with no harmonics generated, and the second
attractor ((II) in Fig.~\ref{fig3}a) corresponds to the steady state solution with the saturated
comb.

The duration of the transition process depends on the nonlinearity parameter $g/\gamma_0$ which
defines the maximal number of photons generated in the comb harmonics. The larger $g$ is, the smaller
is the number. On the other hand, the initial photon number is equal to unity. The growth rate of the
comb is on the order of $\gamma_0$. Therefore, the transient process is longer when $g/\gamma_0$ is
smaller. We have calculated the delay for two cases: $(g/\gamma_0)^{1/2}=5\times10^{-5}$ (estimated
for a small resonator, e.g. \cite{wang11arch}) and $(g/\gamma_0)^{1/2}=10^{-8}$ (estimated for a much
larger resonator). The result of the calculation is shown in Fig.~(\ref{fig3}b).

The behavior of the DC power of light leaving the resonator has a certain peculiarity. The value of
the power exiting the resonator initially decreases and then increases (Fig.~\ref{fig3}b). The
phenomenon can be explained from the stand point of critical coupling \cite{yariv02ptl}. There is no
light at the output of a linear resonator if $\gamma_{0i}=\gamma_{0c}$ and the steady state is
reached; all the pump light is absorbed in the resonator. That is why the amplitude of the exiting
light drops after the pump is on. A number of the pump photons is redistributed between harmonics of
the comb, as the comb is generated. Those harmonics leave the resonator reducing the absorption of
the pumping light. The interference phenomenon resulting in the critical coupling is also
deteriorated since the pump light confined in the corresponding mode changes its amplitude and phase
due to the nonlinear process.

The temporal dependencies shown in Fig.~(\ref{fig1}b) and Fig.~(\ref{fig3}b) can be verified
experimentally if one measures the power of the comb harmonics exiting the resonator. Instead of the
direct measurement of the optical harmonics, the power of the radio frequency (RF) signal generated on a fast photodiode by
the comb  can be measured. We performed such experiments.

We used a calcium fluoride (CaF$_2$) whispering gallery mode (WGM) microresonator with loaded
Q-factor $2.2\times10^9$ (full width at the half maximum of the mode is 90~kHz, and corresponding
ring down time 1.75~$\mu$s). The intrinsic Q-factor of the resonator was $5.5\times10^9$
($\gamma_{0c}=1.5\gamma_{0i}$), which means that the attenuation of light in the modes was primarily
given by the interaction with the evanescent field coupler (a glass prism), and not by the scattering
and loss of the resonator host material. The resonator was pumped using a 1545~nm distributed
feedback (DFB) semiconductor laser, self-injection locked to a selected resonator mode
\cite{liang10ol}. The laser was emitting approximately $6$~mW of power, 30\% of which entered the
resonator.
\begin{figure}[ht]
  \centering
  \includegraphics[width=8.5cm]{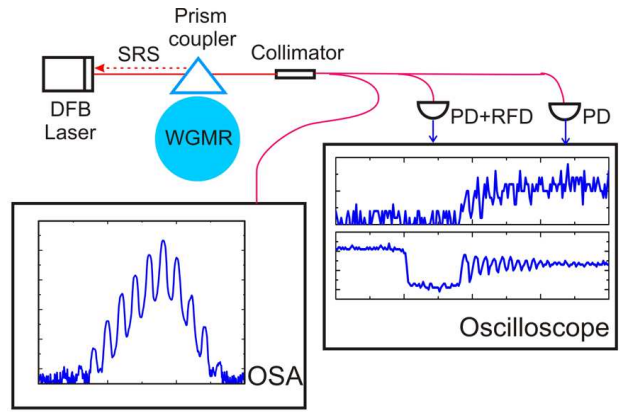}
\caption{ Schematic of the experimental setup. The WGM resonator is pumped with a DFB laser, self
injection locked to the selected mode. The output light is analyzed with an optical spectrum analyzer
(OSA) showing the spectrum of the generated Kerr comb. Part of the light is sent to a fast photodiode
(PD). The photocurrent, modulated with a frequency equal to the comb repetition rate, is directed
to an RF power detector (RFD), and a fast oscilloscope. This signal is proportional to the
convolution of the harmonics of the optical frequency comb. Some of the light is also detected with a
slow photodiode (PD) and the photocurrent from this photodiode is forwarded to another channel of the
same oscilloscope. This signal shows the integral DC power leaving the resonator. Comparing the
signals with the fast oscilloscope we are able to measure the time delay between the generation of
the Kerr comb and the moment the pump light enters the corresponding mode.} \label{fig4}
\end{figure}

\begin{figure}[ht]
  \centering
  \includegraphics[width=8.5cm]{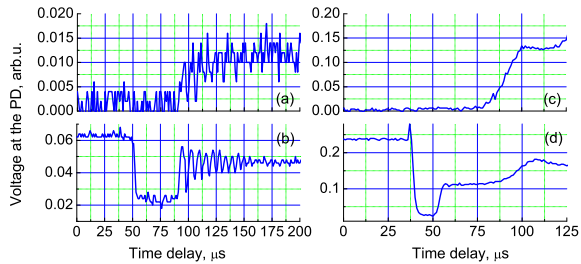}
\caption{ Experimentally observed transient behavior of the Kerr frequency comb measured via
monitoring the power of the DC signal on a slow photodiode generated by the light exiting the
resonator (curves (a) and (c)) as well as the power of the RF signal generated by the comb on a fast
photodiode (curves (b) and (d)). Two modes are considered. The coupling to the mode resulting in the
comb that produces in curves (a) and (b) is much lower compared with the coupling to the mode
responsible for curves (c) and (d). The modes are also characterized with different GVD values.}
\label{fig5}
\end{figure}

We tuned the laser frequency by changing the laser current, and observed the temporal behavior of the
signals at the OSA and the oscilloscope (Fig.~\ref{fig4}). We noticed a significant delay between the
start of the optical pumping of the mode, and the moment of comb generation (Fig.~\ref{fig5}). Some
experimental data  are very similar in shape to the theoretical predictions (Fig.~\ref{fig5}a,b),
while other data are not (Fig.~\ref{fig4}c,d). However the delay is always there. The different
shapes of the signals are measured since we selected different modes in the resonator. It was shown
that the GVD of the modes change depending on the resonator morphology \cite{savchenkov11np} and
depending on the externally pumped mode. In addition to the GVD, the coupling efficiency is also
different for different modes.

It may appear that conditions of the numerical simulation and the experiment described above are
not exactly the same. In the experiment we used a large CaF$_2$ resonator ($6.72$~mm in diameter) with
free spectral range of 10~GHz. The GVD of modes of an ideal spheroidal resonator of this size is much
smaller compared with the value of GVD we utilized in the numerical simulations. Moreover,  GVD is
normal for the fundamental mode sequence of the CaF$_2$ resonator (assuming its ideal spheroidal
shape). Nevertheless, the optical frequency comb observed in the experiment has a spectral shape
similar to the one obtained with the simulation, and the transient processes measured in the
experiment have good correspondence with the theory. This contradiction is resolved if we take into
account the notion that the morphology of a monolithic resonator allows changing the  sign and
value of GVD \cite{savchenkov11np}. In the experiment we did not use the ideal spheroidal resonator or the
fundamental mode sequence. We rather sent the light to a higher-order mode of the resonator. The
observed Kerr frequency comb was generated for light tuned nearly at the top of the WGM resonance
($\omega \simeq \omega_0$). This is possible only if GVD is anomalous and large for a locally
selected mode family, in accordance with the theory of hyper-parametric oscillation
\cite{matsko05pra}. That is why the numerical simulation for a resonator characterized with large
anomalous dispersion is applicable for the description of the experiment performed with a resonator
seemingly having normal dispersion.

To conclude, we have studied the transient regimes of Kerr frequency comb formation in a nonlinear
monolithic optical resonator. We found that the well developed comb generation is delayed by tens of
ring down intervals of the resonator. Kerr combs created in larger and/or less nonlinear resonators
have longer transient period compared with those generated in smaller and/or more nonlinear
resonators. We noted that Kerr combs generated in resonators possessing comparably large anomalous
GVD are always mode-locked. Finally, we experimentally found that there exist mode families with
large anomalous GVD even in WGM resonators made out of a material with normal GVD. We experimentally
validated the results of our numerical simulations with a suitable mode family.




\begin{thebibliography}{99}

\bibitem{kippenberg11s} T. J. Kippenberg, R. Holzwarth, and S. A. Diddams, Science {\bf 332}, 555 (2011).

\bibitem{arcizet09ch} O. Arcizet, A. Schliesser, P. Del'Haye, R. Holzwarth, and T. J. Kippenberg, in {\em Practical Applications of Microresonators in Optics and Photonics,} A. B. Matsko, ed. (CRC Press, Boca Raton, FL, 2009), Chap. 11.

\bibitem{foster11oe} M. A. Foster, J. S. Levy, O. Kuzucu, K. Saha, M. Lipson, and A. L. Gaeta, Opt. Express {\bf 19}, 14233 (2011).

\bibitem{ferdous11np} F. Ferdous, H. Miao, D. E. Leaird, K. Srinivasan, J. Wang, L. Chen, L. Tom Varghese, and A. M. Weiner, "Spectral Line-by-Line Pulse Shaping of an On-Chip Microresonator Frequency Comb," Nature Photonics,  October (2011).

\bibitem{papp11arch} S. B. Papp and S. A. Diddams, "Spectral and temporal characterization of a fused-quartz microresonator optical frequency comb," lanl.arXiv.org$>$physics$>$arXiv:1106.2487v1.

\bibitem{agha09oe} I. H. Agha, Y. Okawachi, and A. L. Gaeta, Opt. Express {\bf 17}, 16209 (2009).

\bibitem{chembo10prl} Y. K. Chembo, D. V. Strekalov, and N. Yu, Phys. Rev. Lett. {\bf 104}, 103902 (2010).

\bibitem{chembo10pra} Y. K. Chembo and N. Yu, Phys. Rev. A {\bf 82}, 033801 (2010).

\bibitem{matsko11ol} A. B. Matsko, A. A. Savchenkov, W. Liang, V. S. Ilchenko, D. Seidel, and L. Maleki, Opt. Lett. {\bf 36}, 2845 (2011).

\bibitem{delhaye11prl} P. Del'Haye, T. Herr, E. Gavartin, M. L. Gorodetsky, R. Holzwarth, and T. J. Kippenberg, Phys. Rev. Lett. {\bf 107}, 063901 (2011).

\bibitem{savchenkov11np} A. A. Savchenkov, A. B. Matsko, W. Liang, V. S. Ilchenko, D. Seidel, and L. Maleki, Nature Photonics {\bf 5}, 293 (2011).

\bibitem{matsko05pra} A. B. Matsko, A. A. Savchenkov, D. Strekalov, V. S. Ilchenko, and L. Maleki,  Phys. Rev. A {\bf 71}, 033804 (2005).

\bibitem{yariv02ptl} A. Yariv, IEEE Photon. Tech. Lett. {\bf 14}, 483 (2002).

\bibitem{maleki10ifcs} L. Maleki, V. S. Ilchenko, A. A. Savchenkov, W. Liang, D. Seidel, and A. B. Matsko, Proc. IEEE Int. Freq. Cont. Symp. {\bf 1}, 558 (2010).

\bibitem{wang11arch} C. Y. Wang, T. Herr, P. Del'Haye, A. Schliesser, J. Hofer, R. Holzwarth, T. W. H\"ansch, N. Picque, and T. J. Kippenberg, "Mid-Infrared Optical Frequency Combs based on Crystalline Microresonators," lanl.arXiv.org$>$physics$>$arXiv:1109.2716.

\bibitem{liang10ol} W. Liang, V. S. Ilchenko, A. A. Savchenkov, A. B. Matsko, D. Seidel, and L. Maleki, Opt. Lett. {\bf 35}, 2822 (2010).


\end{thebibliography}
\end{document}